\documentstyle[12pt]{article}
\def\a{\alpha}
\def\b{\beta}
\def\db{{\dot{\beta}}}
\def\t{\theta}
\def\bt{{\bar{\theta}}}
\def\yb{{\bar{y}}}
\def\Db{{\bar{D}}}
\def\Tr{{\rm Tr}}
\def\dis{\displaystyle}
\def\le{\left(}
\def\ri{\right)}
\def\da{{\dot{\alpha}}}
\def\no{\nonumber}
\def\del{\delta}
\def\bW{\bar{W}}
\def\G{\Gamma}
\def\rar{\rightarrow}
\def\fra1g2{\frac{1}{g^2}}
\def\dg{{\dagger}}
\def\Vt{\tilde{V}}
\def\Kt{\tilde{K}}
\def\SQa{\sqrt{\tilde{\a}}}
\def\e{\epsilon}
\def\sm{{\sigma_m}}
\def\f12{\frac{1}{2}}

\begin{document}
\begin{titlepage}
\flushright{SISSA/73/00/EP}\\
\vskip 2cm
\begin{center}
{\Large \bf The solution to Slavnov--Taylor identities \\
\vspace{3mm} in D4 N=1 SYM} 
\vskip 1cm
 Igor Kondrashuk\footnote{E-mail: ikond@sissa.it, 
on leave of absence from LNP, JINR, Dubna, Russia}
\vskip 5mm
{\it SISSA -- ISAS  and INFN, Sezione di Trieste, \\
 Via Beirut 2-4, I-34013, Trieste, Italy}
\end{center}
\vskip 20mm
\begin{abstract}
D4 N=1 SYM with an arbitrary chiral background superfield 
as the gauge coupling is considered. The solution to Slavnov--Taylor 
identities has been given. It has been shown that the solution is unique and allows  
us to restrict the gauge part of the effective action. 
Under the effective action in this paper we mean the 1PI diagram generator.    
\end{abstract}
\vskip 1cm
\begin{center}
\end{center}
\end{titlepage}

One of the ways to break supersymmetry is to introduce  interactions 
with background superfields that are space-time independent into the 
supersymmetric theory. If we do not include these background superfields 
in the supersymmetry transformation at the component level we have 
breaking of supersymmetry. This is so-called softly broken supersymmetry.
The relation between the theory with softly broken supersymmetry and 
its rigid counterpart has been studied in Refs. (\cite{Yam}-\cite{Giudice-2}).
The investigation has been performed for singular parts 
of the effective actions of softly broken and rigid theories. 
Since the only modification of the classical action from the rigid 
case to the softly broken case is a replacement of coupling constants 
of the rigid theory with background superfields,
the relation is simple and can be reduced to substitutions of  
these superfields into renormalization constants of the rigid theory 
instead of the rigid theory couplings \cite{Jones,AKK}.      
Later, a relation between full correlators of softly broken
and unbroken SUSY quantum mechanics has been found \cite{IKQM}.
Nonperturbative results for the terms of the effective action 
which correspond to the case when chiral derivatives do not act on  
background superfields have been derived in Ref. \cite{Ratt}. 

The renormalization of the soft theory has been made 
on the basis of supergraph technique in the Ref.\cite{AKK}, and 
at the level of Slavnov--Taylor identities the renormalization procedure 
for these theories has been performed more recently in \cite{IKst2m} for the 
case of an arbitrary chiral superfield as the gauge coupling.   
However, the renormalization procedure deals with infinities that must be removed 
from the theory. Actually, Slavnov--Taylor (ST) identities  contain much more 
information since they restrict the total effective action $\G.$ They could be used 
to extract information about finite parts of proper correlators. This information 
can be useful in order to calculate process amplitudes in the Minimal 
Supersymmetric Standard Model within superfield formalism. In this letter we propose 
the way to find the solution to ST identities and show that the solution 
is unique.   

The notation used for the D4 supersymmetry and for the classical 
action $S$ of the theory with softly broken supersymmetry are the same as those
have been used in our previous paper \cite{IKst2m}. In this work  
we concentrate on the pure gauge supersymmetric theory. 

The gauge fixing condition \cite{IKst2m} is taken as 
\begin{eqnarray}
D^2\frac{V(x,\t,\bt)}{\sqrt{\tilde{\a}}}  = \bar{f}(\yb,\bt),  ~~~  
\Db^2 \frac{V(x,\t,\bt)}{\sqrt{\tilde{\a}}} = f(y,\t), \label{mod} 
\end{eqnarray}
with an arbitrary background gauge fixing superfield $\tilde{\a}.$ We 
have shown in \cite{IKst2m} that this modification of the 
gauge fixing condition is necessary and important for the renormalization 
procedure in the softly broken SYM. 
  
With the modification (\ref{mod}) the total gauge part of the classical action  
is  
\begin{eqnarray}
& S_{\rm{gauge}} = \dis{\int d^4y d^2\t~S\frac{1}{2^7}\Tr~ W_\a W^\a  +  
\int d^4\yb d^2\bt~\bar{S}\frac{1}{2^7}\Tr~ \bW^\da \bW_\da } \no \\
& + \dis{\int d^4x d^2\t d^2\bt ~\frac{1}{16}
\Tr~\le\Db^2 \frac{V}{\sqrt{\tilde{\a}}}\ri\le D^2\frac{V}{\sqrt{\tilde{\a}}}\ri}
  \label{SSgauge} \\ 
& + \dis{\int d^4y d^2\t~\frac{i}{2}\Tr~  b~\Db^2 
\le\frac{\del_{\bar{c},c}V}{\sqrt{\tilde{\a}}}\ri  + 
\int d^4\yb d^2\bt~\frac{i}{2}\Tr~  \bar{b}~D^2 
\le\frac{\del_{\bar{c},c}V}{\sqrt{\tilde{\a}}}\ri.} \no
\end{eqnarray}
Here $S$ is an arbitrary chiral superfield (in general, $x$-dependent). For 
example, in case of the softly broken supersymmetry we have 
\begin{eqnarray*}
\dis{S = \fra1g2\le 1 - 2m_A\t^2 \ri}. 
\end{eqnarray*}
In the softly broken case the path integral describing the quantum 
soft theory is defined as 
\begin{eqnarray}
& \dis{Z[J,\eta,\bar{\eta},\rho,\bar{\rho},K,L,\bar{L}] = 
\int dV~dc~d\bar{c}~db~d\bar{b}~\exp i}\left[\dis{S_{\rm{gauge}}} 
\right.  \label{pathS}\\
& \left. + \dis{2~\Tr\le JV + i\eta c + i\bar{\eta}\bar{c} + i\rho b + 
i\bar{\rho}\bar{b}\ri + 
2~\Tr\le iK\del_{\bar{c},c}V + Lc^2 + \bar{L}\bar{c}^2 \ri}\right]. \no
\end{eqnarray}
The third term in the brackets is BRST invariant since the external 
superfields $K$ and $L$ are BRST invariant by definition. All fields
in the path integral are in the adjoint representation of the gauge group.
For the sake of brevity we omit the symbol of integration in the 
terms with external sources, keeping in mind that it is the full superspace 
measure for vector superfields and the chiral measure for chiral superfields.

Having shifted the antighost superfields $b$ and $\bar{b}$ by arbitrary 
chiral and antichiral superfields respectively, or, having made the  
change of variables in the path integral (\ref{pathS}) which are the 
BRST transformations  of the total gauge action 
(\ref{SSgauge}), we get ghost equations \cite{Piguet} 
\begin{eqnarray}
\frac{\del \G}{\del\bar{b}}  - \frac{1}{4}D^2\frac{1}{\sqrt{\tilde{\a}}}
\frac{\del \G}{\del K} = 0,  ~~~
\frac{\del \G}{\del b} - \frac{1}{4}\Db^2\frac{1}{\sqrt{\tilde{\a}}}
\frac{\del \G}{\del K} = 0 \label{ghostEs}
\end{eqnarray}
and ST identities   
\begin{eqnarray}
& \Tr\left[\dis{\frac{\del \G}{\del V}\frac{\del \G}{\del K} 
 - i \frac{\del \G}{\del c}\frac{\del \G}{\del L}  
+ i \frac{\del \G}{\del \bar{c}}\frac{\del \G}{\del \bar{L}} 
- \frac{\del \G}{\del b}\le\frac{1}{32}\Db^2D^2\frac{V}{\sqrt{\tilde{\a}}}\ri} 
\right. \label{STs}\\ 
& \left. - \dis{\frac{\del \G}{\del \bar{b}}\le\frac{1}{32}D^2\Db^2\frac{V}
{\sqrt{\tilde{\a}}}\ri}\right]=0, \no
\end{eqnarray}
respectively. In the Ref. \cite{IKst2m} it has been shown that the  
ghost equations (\ref{ghostEs}) and the ST identities (\ref{STs}) 
restrict the singular part of the effective action of the SYM theory 
with softly broken supersymmetry which corresponds to the theory with 
classical action (\ref{SSgauge}) to  
\begin{eqnarray}
& \G_{\rm{sing}} = \dis{\int d^4y d^2\t~S\frac{1}{2^7}~
\tilde{z}_S 
\Tr~ W_\a\le\frac{V}{\tilde{z}_1}\ri W^\a\le\frac{V}{\tilde{z}_1}\ri  + 
{\rm H.c.}}  \no\\
& + \dis{\int  d^4x d^2\t d^2\bt~\Tr\frac{1}{32}\frac{V}
{\sqrt{\tilde{\a}}} \le D^2\Db^2 + \Db^2D^2\ri 
\frac{V}{\sqrt{\tilde{\a}}}}  \label{GsingS}\\
& + \dis{\int  d^4x d^2\t d^2\bt~2i~\Tr\le 
(b + \bar{b})\frac{\tilde{z}_1}{\sqrt{\tilde{\a}}} + K\tilde{z}_1\ri~\left[
\del_{\bar{c},c}\le\frac{V}{\tilde{z}_1}\ri\right],} \no
\end{eqnarray} 
where the singular superfields $\tilde{z}_1$ and  $\tilde{z}_S$ are related 
to the renormalization constants of the rigid theory in a simple way
\cite{IKst2m}, 
\begin{eqnarray*}
& \dis{\tilde{z}_1\le S,\bar{S},\sqrt{\tilde{\a}}\ri = 
z_1\le g^2 \rar\le\frac{S + \bar{S}}{2}\ri ^{-1} , 
\sqrt{\a} \rar \sqrt{\tilde{\a}} \ri},  \label{rule}\\
& \dis{\tilde{z}_S(S) =  z_{g^2}\le\fra1g2 \rar S\ri. } \no
\end{eqnarray*}

Here we show that the total effective action can be restricted in an  
analogous way. We go alone the line of our previous paper \cite{IKst2m}.
The ghost equations  (\ref{ghostEs}) restrict the dependence of $\G$ on the antighost 
superfields and on the external source $K$ to an arbitrary dependence on their 
combination 
\begin{eqnarray}
\le b + \bar{b}\ri\frac{1}{\sqrt{\tilde{\a}}}  + K. \label{comb}
\end{eqnarray}
We can present this dependence of the effective action on the external source 
$K$ in terms of a series  
\begin{eqnarray}
\G = {\cal{F}}_0 + \sum_{n=1}\int d^8z_1d^8z_2\dots d^8z_n {\cal{F}}_n
\le z_1, z_2,\dots, z_n\ri K(z_1)K(z_2)\dots K(z_n). \label{ser} 
\end{eqnarray}
According to (\ref{comb}) we should write 
$\le b(z_i) + \bar{b}(z_i)\ri{/}\sqrt{\tilde{\a}(z_i)}  + K(z_i)$
instead of $K(z_i),$ but we do not do it for the sake of brevity.  
The coefficient functions of this expansion are in their turn 
functionals of other superfields, 
\begin{eqnarray*}
{\cal{F}}_n = {\cal{F}}_n\left[V,c,\bar{c}, L, \bar{L}\right],    
\end{eqnarray*}
whose coefficient functions are  ghost-antighost-vector 
correlators. ${\cal{F}}_0$ is a $K$-independent part of the 
effective action.

Our purpose is to restrict the expansion (\ref{ser}) by using 
ST identities (\ref{STs}). At the moment for simplicity we suggest 
that the coefficient functions ${\cal{F}}_n$ for $n>1$ do not depend on 
the external sources $L,\bar{L}$ (in what follows we argue this conjecture). 
The total degree of the ghost superfields $c$ and $\bar{c}$ 
in ${\cal{F}}_n$ must be equal to $n$ since each proper supergraph 
contains equal number of ghost and antighost superfields among its 
external legs.  

To start we consider ${\cal{F}}_1(z_1)$ coefficient function in the 
expansion (\ref{ser}). The corresponding term in (\ref{ser}) is   
\begin{eqnarray}
\dis{\int  d^8z~d^8z'~2i~\Tr\le \le b(z) +\bar{b}(z)\ri\frac{1}{\sqrt{\tilde{\a}(z)}}
 + K(z)\ri~G^{(2)}_{[S,\bar{S},\sqrt{\tilde{\a}}]}(z-z')\le c(z') + \bar{c}(z')\ri}, 
\label{K1}
\end{eqnarray}
where $G^{(2)}_{[S,\bar{S},\sqrt{\tilde{\a}}]}(z-z')$ is a 2-point ghost 
proper correlator. It is functional of external background superfields 
$S,\bar{S},\sqrt{\tilde{\a}}.$ It is a Hermitian kernel of the above integral,      
\begin{eqnarray*}
\le G^{(2)}_{[S,\bar{S},\sqrt{\tilde{\a}}]} \ri ^{\dg} = 
 G^{(2)}_{[S,\bar{S},\sqrt{\tilde{\a}}]}.
\end{eqnarray*} 
We make the change of variables in the effective active action $\G,$ 
\begin{eqnarray}
& \dis{K(z) = \int d^8z'~\tilde{K}(z')~{}^{(-1)}
G^{(2)}_{[S,\bar{S},\sqrt{\tilde{\a}}]}(z-z')},
\label{conv} \\
& \dis{V(z) = \int d^8z'~\tilde{V}(z')~G^{(2)}_{[S,\bar{S},
\sqrt{\tilde{\a}}]}(z-z')}
, \no 
\end{eqnarray}
where ${}^{(-1)}G^{(2)}_{[S,\bar{S},\sqrt{\tilde{\a}}]}(z-z')$ is a 
2-point connected ghost correlator, 
\begin{eqnarray*}
\int d^8z'~G^{(2)}_{[S,\bar{S},\sqrt{\tilde{\a}}]}(z_1 - z')~
{}^{(-1)}G^{(2)}_{[S,\bar{S},\sqrt{\tilde{\a}}]}(z_2 - z') = 
\del^{(8)}(z_2 - z_1).
\end{eqnarray*}
In terms of new variables, the effective action 
\begin{eqnarray}
& \dis{\G\left[V,c,\bar{c},b,\bar{b},K,L,\bar{L}\right] = 
\G\left[V(\Vt),c,\bar{c},b,\bar{b},K(\Kt),L,\bar{L}\right]} \no\\
& \dis{\equiv 
\tilde{\G}\left[\Vt,c,\bar{c},b,\bar{b},\Kt,L,\bar{L}\right]}, 
\label{changeSf}
\end{eqnarray}
satisfies to modified ST identities 
\begin{eqnarray}
& \Tr\left[\dis{\frac{\del \tilde{\G}}{\del \Vt}\frac{\del \tilde{\G}}{\del \Kt} 
 - i \frac{\del \tilde{\G}}{\del c}\frac{\del \tilde{\G}}{\del L}  
+ i \frac{\del \tilde{\G}}{\del \bar{c}}\frac{\del\tilde{\G}}{\del \bar{L}} 
- \frac{\del\tilde{\G}}{\del b}\le\frac{1}{32}\Db^2D^2
\frac{\Vt\star G^{(2)}_{[S,\bar{S},\sqrt{\tilde{\a}}]}}
{\sqrt{\tilde{\a}}}\ri} \right. 
\label{STsmf}\\ 
& \left. - \dis{\frac{\del \tilde{\G}}{\del \bar{b}}
\le\frac{1}{32}D^2\Db^2
\frac{\Vt\star G^{(2)}_{[S,\bar{S},\sqrt{\tilde{\a}}]}}
{\sqrt{\tilde{\a}}}\ri}\right]=0, \no
\end{eqnarray}
where a new notation is introduced for the convolutions (\ref{conv}) 
\begin{eqnarray*}
K = \Kt \star{}^{(-1)} G^{(2)}_{[S,\bar{S},\sqrt{\tilde{\a}}]},~~~
V = \Vt \star ~G^{(2)}_{[S,\bar{S},\sqrt{\tilde{\a}}]}. 
\end{eqnarray*}
Having made the change (\ref{changeSf}), we can present (\ref{K1})
as 
\begin{eqnarray*}
& \dis{\int  d^8z~d^8z'~2i~\Tr \le b(z) +\bar{b}(z)\ri\frac{1}{\sqrt{\tilde{\a}(z)}}
~G^{(2)}_{[S,\bar{S},\sqrt{\tilde{\a}}]}(z-z')\le c(z') + \bar{c}(z')\ri} \\ 
& + \dis{\int  d^8z~2i~\Tr~\Kt(z)~\le c(z) + \bar{c}(z)\ri}.
\end{eqnarray*}
Hence, the first term in the ST identities (\ref{STsmf}) restricts 
$\Vt^2$ term in $\Kt$-independent part of $\tilde{\G}$ to the form  
\begin{eqnarray}
\dis{~\int d^8z~f\left[S,\bar{S},\SQa\right]
~\le D_\a\Vt\ri\le\Db^2D^\a\Vt\ri + {\rm H.c.}},  \no
\end{eqnarray}
where $f\left[S,\bar{S},\SQa\right]$ is a chiral functional of the external 
background superfields. We work in terms of the perturbation 
theory. Hence, we can apply the no-renormalization theorem 
that states the absence of perturbative corrections to the 
superpotential. This means that the superpotential of (\ref{pathS})
\begin{eqnarray*}
\dis{\int d^4y d^2\t~2\Tr~Lc^2 + 
\int d^4\yb d^2\bt~2\Tr~\bar{L}\bar{c}^2} 
\end{eqnarray*}
remains unchanged in the effective action. The first term in 
ST identities also can be expanded in terms of $\Kt,$ 
\begin{eqnarray*}
\dis{\frac{\del \tilde{\G}}{\del \Vt}\frac{\del \tilde{\G}}{\del \Kt} = {\cal{M}}_0
+ \sum_{n=1}\int d^8z_1d^8z_2\dots d^8z_n~{\cal{M}}_n
\le z_1, z_2,\dots, z_n\ri \Kt(z_1)\Kt(z_2)\dots \Kt(z_n)},  
\end{eqnarray*}
where ${\cal{M}}_0$ is $\Kt$-independent part of 
$\dis{\frac{\del \tilde{\G}}{\del \Vt}\frac{\del \tilde{\G}}{\del \Kt}}.$
We can consider it as a result of infinitesimal transformation 
$\tilde{\G}$ in which instead of $\Vt$ we have substituted  
\begin{eqnarray}
\Vt \rar \Vt + \frac{\del \tilde{\G}}{\del \Kt}. \label{subs}
\end{eqnarray}
Hence, the term of the order $\Vt$ in  ${\cal{F}}_1$ is fixed completely by the  
$\Vt^0$ term in ${\cal{F}}_1$ since the only contribution into the  $\Vt^0$ term
in ${\cal{M}}_1$ comes from this $\Vt$ term in  ${\cal{F}}_1$. Indeed, 
another possible contribution into the $\Vt^0$ term of ${\cal{M}}_1$ could 
go from ${\cal{F}}_2$ substituted in accordance with (\ref{subs}) into  ${\cal{F}}_0$
but it is with necessity of the order of $\Vt$ at least. Hence, the term of the
order of  $\Vt$ in  ${\cal{F}}_1$ is the term of the order of $\Vt$ in 
$\del_{\bar{c},c}\Vt.$ All the terms in ${\cal{F}}_1$ and ${\cal{F}}_0$ of 
higher orders in $\Vt$ are fixed by themselves in an iterative way. 
The only solution to ${\cal{F}}_1$ is $\del_{\bar{c},c}\Vt.$ Starting from the 
fourth degree of $\Vt$ higher order BRST invariant contributions like  
\begin{eqnarray*}
\int d^4y d^2\t~f_2\left[S,\bar{S},\SQa\right]~\Tr~W_\a(\Vt) 
W^\a(\Vt) W_\b(\Vt) W^\b(\Vt) 
\end{eqnarray*}
into ${\cal{F}}_0$ are allowed. Here $f_2\left[S,\bar{S},\SQa\right]$ 
is a chiral functional of external 
background superfields. The following notation is used for brevity 
\begin{eqnarray*} 
\dis{W^\a\le V\ri \equiv \Db^2 \le e^{-V}D^\a e^V \ri.} 
\end{eqnarray*}   
All coefficient functions ${\cal{F}}_n$ with $n > 1$ are equal to zero. 
Indeed,  ${\cal{F}}_2$ contributes into ${\cal{M}}_1$ but we do not have 
anything that can compensate this contribution by ghost transformations 
induced by the second and third terms in the modified ST identities 
(\ref{STsmf}). Hence,  ${\cal{F}}_2 = 0.$ If we consider ${\cal{F}}_3$ it 
contributes into  ${\cal{M}}_2$ and, in general, could be compensated 
by ghost transformations in ${\cal{F}}_2$. But ${\cal{F}}_2$ is zero,
hence, ${\cal{F}}_3$ is also zero. We can repeat the former argument 
for all higher numbers $n$ of ${\cal{F}}_n.$ 

It can be also shown that there is no room for the dependence 
of ${\cal{F}}_n$ on the external superfields $L$ and $\bar{L}.$
Indeed, the first degree of $(L + \bar{L})$ could be written 
in the ${\cal{F}}_1$ term, for example. The corresponding 
contribution into ${\cal{M}}_1$ due to substitution (\ref{subs}) would be 
proportional to $(L + \bar{L})$ while a contribution due to variations 
of the ghost fields in ${\cal{F}}_1$ which are proportional 
to $(L + \bar{L})$ goes into  ${\cal{M}}_2,$ and, hence, they can not 
compensate each other. Hence, the first degree is equal to zero. 
The same is true for all degrees of $(L + \bar{L}).$   

Hence, the effective action can be presented as 
\begin{eqnarray}
& \dis{\tilde{\G}  = \int d^4y d^2\t~f\left[S,\bar{S},\SQa\right]~
\Tr~ W_\a(\Vt) W^\a(\Vt)  + {\rm H.c.}}  \no\\
& + \dis{\int d^4y d^2\t~f_2\left[S,\bar{S},\SQa\right]~
\Tr~W_\a(\Vt) W^\a(\Vt) W_\b(\Vt) W^\b(\Vt) 
  + {\rm H.c.} + \dots} \no\\
& + \dis{\int  d^8z~\frac{1}{32}\Tr
\frac{\Vt\star G^{(2)}_{[S,\bar{S},\SQa]}}{\SQa}
\le D^2\Db^2 + \Db^2D^2\ri  
\frac{\Vt\star G^{(2)}_{[S,\bar{S},\SQa]}}{\SQa}} \label{auxfull} \\
& + \dis{\int  d^8z~d^8z'~2i~\Tr \le b(z) +\bar{b}(z)\ri\frac{1}{\SQa}
~G^{(2)}_{[S,\bar{S},\SQa]}(z-z')~\del_{\bar{c},c}\Vt(z')} \no\\ 
& + \dis{\int  d^8z~2i~\Tr~\Kt(z)~\del_{\bar{c},c}\Vt(z)}. \no
\end{eqnarray}
where $\dots$ denote the BRST (gauge) invariant terms of higher orders 
in $W_\a\le\Vt\ri.$
Now we should go back to the initial variables $V$ and $K$, that is, we should 
made the change of variables in $\tilde{\G}$ reversed to (\ref{changeSf}). 
\begin{eqnarray}
& \dis{\G  = \int d^4y d^2\t~f\left[S,\bar{S},\SQa\right]~
\Tr~ W_\a\le V \star{}^{(-1)} G^{(2)}_{[S,\bar{S},\SQa]}\ri 
W^\a\le V \star {}^{(-1)}G^{(2)}_{[S,\bar{S},\SQa]}\ri  
+ {\rm H.c.}}  \no\\
& + \dis{\int d^4y d^2\t~f_2\left[S,\bar{S},\SQa\right]~
\Tr~}\le \dis{W_\a\le V \star{}^{(-1)} G^{(2)}_{[S,\bar{S},\SQa]}\ri
W^\a\le V \star {}^{(-1)}G^{(2)}_{[S,\bar{S},\SQa]}\ri}  \right. \no\\ 
& \left. \dis{W_\b \le V \star {}^{(-1)}G^{(2)}_{[S,\bar{S},\SQa]}\ri
W^\b \le V \star {}^{(-1)}G^{(2)}_{[S,\bar{S},\SQa]}\ri} \ri 
  + {\rm H.c.} + \dots \no\\
& + \dis{\int  d^8z~\frac{1}{32}\Tr
\frac{V}{\SQa}\le D^2\Db^2 + \Db^2D^2\ri\frac{V}{\SQa}} \label{Gfull}\\
& + \dis{\int  d^8z~d^8z'~2i~\Tr \le b(z) +\bar{b}(z)\ri\frac{1}{\sqrt{\tilde{\a}(z)}}
~G^{(2)}_{[S,\bar{S},\sqrt{\tilde{\a}}]}(z-z')~\del_{\bar{c},c}
\le V \star{}^{(-1)} G^{(2)}_{[S,\bar{S},\SQa]}\ri (z')  } \no\\ 
& + \dis{\int  d^8z~2i~\Tr~\le K\star G^{(2)}_{[S,\bar{S},\SQa]}\ri (z)
~\del_{\bar{c},c}\le V \star{}^{(-1)} G^{(2)}_{[S,\bar{S},\SQa]}\ri (z)}. \no
\end{eqnarray}
As usual, forth and fifth terms in the modified ST identities
together with the ghost equation do not allow any correction to the
gauge fixing term \cite{IKst2m}.  

All the derivations proposed in this paper have a sense only if 
we have fixed a gauge invariant regularization 
and defined a renormalization scheme to remove infinities. In this work 
we implied the DRED scheme \cite{DRED} that is the only practical 
regulator in order to be able to calculate higher order effects in any 
supersymmetric theory including MSSM. In this case ST    
identities (\ref{STsmf}) do not forbid a new gauge invariant term \cite{GMZ} 
of the effective action $\tilde{\G}$ (\ref{auxfull}) 
\begin{eqnarray*} 
\int  d^4x d^2\t d^2\bt~g_{mn}^{(\e)}~\Tr\G_m\G_n
\end{eqnarray*} 
where $g_{mn}^{(\e)}$ is the metric in the $2\e$ compactified dimensions
and $\G_m$ is the superfield gauge connection defined by 
\begin{eqnarray*}
\G_m = \f12 \sm^{\a\db}\Db_{\db}\le e^{-V}D_{\a}e^V \ri. 
\end{eqnarray*}
This term generates so-called $\e$ scalar masses \cite{cinque} in the course of the 
renormalization procedure. Indeed, one can see that at the component level 
the $\t^2\bt^2$ component of $\tilde{z}_1$ that is the singular term of 
$G^{(2)}_{[S,\bar{S},\SQa]}$  produces $\e$ scalar masses 
when we are replacing $\Vt$ with $V \star{}^{(-1)} G^{(2)}_{[S,\bar{S},\SQa]}$ 
in $\tilde{\G}$ to obtain the effective action $\G$ (\ref{Gfull}).
Even if initially the $\e$ scalar masses are equal to zero, this 
condition is unstable under renormalizations and, hence, such a 
counterterm must be added. As it has been found in \cite{cinque}
$\e$ scalar masses dependence of the two-loop $\b$ functions can be completely 
removed by a slight modification of the DRED scheme to the $\overline{\rm DR'}$    
scheme. The way how to generalize this scheme to all orders of the 
perturbation theory has been proposed in Ref.\cite{Giudice-2}.    
However, based on the explicit presence of this 
contribution at the two-loop level in the component formalism 
\cite{cinque}, it has been stated in \cite{Jones11} that  
the contribution of the $\e$ scalar mass renormalization 
should be taken into account in the physical soft scalar mass 
$\b$ functions. It is possible to determine this contribution at 
all orders of the perturbation theory by requiring the existence of a
set of renormalization group invariant relations between soft couplings and 
masses as it has been done in Ref. \cite{Jones11} for the $\overline{\rm DR'}$
scheme  and further developed in Refs. \cite{Jones12,Jones13} to other schemes.     

In summary, we have shown the solution to ST identities can be 
parametrized in the form (\ref{Gfull}). We have demonstrated  
that there are not quantum contributions into the effective action 
containing more than two ghost fields. This solution has been obtained without using 
the background field technique. By construction one can see that 
the solution is unique. This solution can have an application 
to calculations of process amplitudes in the MSSM containing 
components of gauge superfields as asymptotic states by using superfield
formalism without going to components.  
\vskip 3mm 
\noindent {\large{\bf{Acknowledgements}}} 
\vskip 3mm
I am grateful to Antonio Masiero and Giulio Bonelli for useful discussions. 
This work is supported by INFN.

\end{document}